# Revisiting the Mousetrap Experiment: Not Just About Nuclear Chain Reactions

By Ugo Bardi and Ilaria Perissi
Dipartimento di Chimica, Università di Firenze, Italy
ugo.bardi@unifi.it, ilariaperissi@gmail.com

Abstract. We present here quantitative measurements on a classic experiment proposed for the first time in 1947 to illustrate the phenomenon of a chain reaction in nuclear fission. The experiment involves a number of mousetraps loaded with solid balls. Once one trap is made to snap, it releases two balls that cause other traps to snap more traps and the result is a chain reaction. The experiment has been popular as a scientific demonstration, but it doesn't seem that quantitative data were ever reported about it, nor that it was described using a model. We set out to do that and we can report for the first time that the mousetrap experiment can be fitted by a simple dynamic model derived from the well known Lotka-Volterra one. We also discuss the significance of this experiment beyond nuclear chain reactions, as providing insight in a variety of fields involving complex, adaptive systems.

## 1 Introduction

The "Mousetrap Experiment" consists in loading 25-100 mousetraps with solid balls. Triggering one of the traps results in the release of two balls, which then trigger other traps, generating a rapid explosion of flying balls that subsides when most of the traps have been triggered. It was proposed in1947 by Richard Sutton (1) as a mechanical simulation of the chain reaction that occurs in fissile nuclei, such as the 235 isotope of Uranium. The trap represents a nucleus, while the balls represent the neutrons that trigger fission in other nuclei. The simulation creates a chain reaction, as it happens in nuclear warheads. The experiment was widely popularized in the 1950s and it was shown in Walt Disney's movie, "*Our Friend, The Atom*" (1957).

Today, tens, perhaps hundreds, of examples of the mousetrap experiment can be found over the Web. But it doesn't appear that anyone ever quantified the parameters of the reaction and tried to reproduce it by a mathematical model. This is what we set out to do by measurement and by means a simple system dynamic model, the "Single Cycle Lotka-Volterra" model that we developed in earlier papers to describe the phenomenon of overexploitation in economic systems (2), (3). We found that the model can describe the dynamics of the mousetrap experiment. We believe that this approach adds interest to an experiment that has great value for the dissemination of dynamic concepts not just in nuclear physics, but also in fields such as ecology, economics and memetics.

## 2. Experimental

We followed two procedures to obtain data for the mousetrap experiment. One was to look for video clips on the Web, the other was to set up a specific experiment in Florence, Italy (we call it, "The Florence Experiment.") We found only a few experiments on the Web where the experiment was performed in such a way that it could be quantified. In most cases, the traps are not fixed to the supporting plate, so that the result is that balls and traps fly in the air together during the chain reaction. Some experiments are better thought out and executed, such as the Berkeley Mousetrap Experiment (4), (5). Another well managed and executed experiment is the one reported by Dalton Nuclear (6).

The Florence experimental set-up was built using 50 commercial mousetraps, each of 50x95 mm. The traps were set on five rows of 10 traps each, spaced about 5 mm from each other and covering a square of 55x55 cm. They were glued to a soft underlying material in order to avoid vibrations that could cause the traps to trigger each other. The square was enclosed in a cubic plexiglass enclosure of 62 cm x 62 cm. Each trap was loaded with two wooden balls of 30 mm diameter. The sensitivity of the triggering mechanism was enhanced by fixing 25 mm rigid cardboard disks to the original metal trigger. In these conditions, we found that the chain reaction was completed in ca. 2-3 seconds, leaving most of the traps triggered. As reported by other authors (7), setting 50 traps requires a certain manual dexterity that the authors of this paper acquired by trial and error, suffering only minor damage to their fingers in the process.

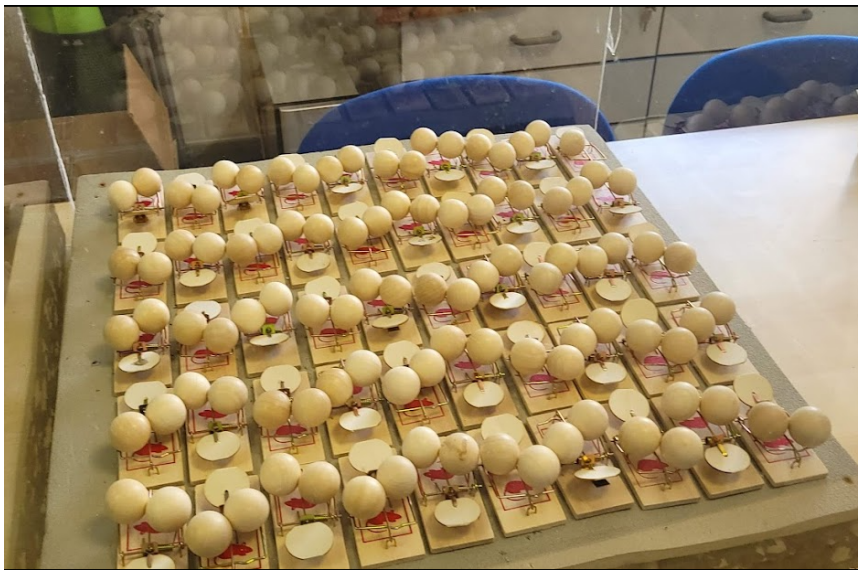

Fig. 1. The Florence mousetrap experiment

In all cases, we determined the parameters of the chain reaction using video recordings to count the flying balls and the triggered traps. For the Florence experiment, we used commercial Samsung cell phones in slow motion of ½ or ¼ of the real speed. For the Web experiments we used the video clips available. In all cases, we used the site www.watchframebyframe.com to count the flying balls as a function of fixed intervals of time. Typically, we counted every 1/10th of a second.

Counting the balls in this way involved a certain degree of uncertainty related to detecting moving balls close to the traps and to reflections of the balls on the plexiglas walls enclosing the experimental volume. Nevertheless, repeated tests on the same experiments showed that the uncertainty was at most one-two balls and it didn't prevent the determination of the kinetics of the experiment.

**3. Results.**

3.1 *The Berkeley experiment.* This test involved 49 mousetraps, each one loaded with two plastic balls ("superballs"). Unlike most set-ups, the traps were not triggered by a spring, but by a solenoid operated by a metal strip. The chain reaction lasted about 4 seconds. From the video, it was possible to count the flying balls, but not the number of the triggered traps as a function of time. The results are shown in Fig. 2. The experimental points have been fitted with the derivative of a logistic function. Note the "bell shaped" form of the curve: it is a typical feature of these mousetraps experiments.

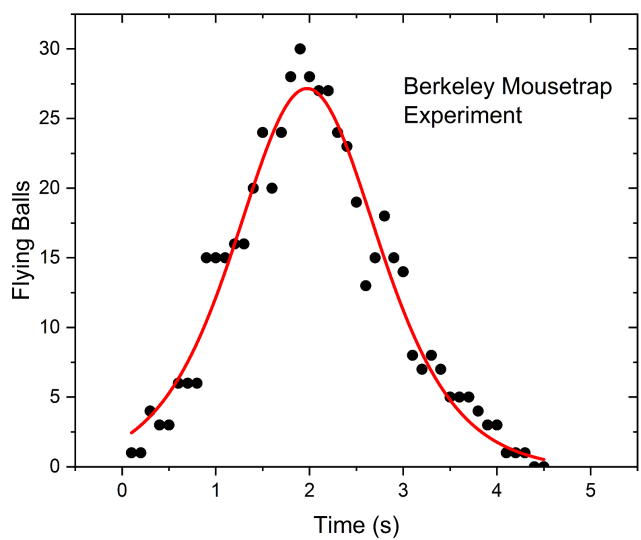

Fig. 2. The results of the Berkeley Mousetrap experiment

3.2 *The Dalton Nuclear experiment.* In the description, the experiment is said to have involved 200 traps loaded with one ball each. The images, however, showed that only 110 traps (and as many balls) were used. The experiment was shown in slow motion and the chain reaction as shown lasted for 90 seconds. Unfortunately, the authors do not report the motion rate they used. Assuming that each frame corresponds to 0.25 sec, the whole reaction may have lasted around 3 seconds. Although the time scale is uncertain, the shape of the curve for the flying ballsl is again "bell-shaped." Also in this case, it was not possible to determine the number of triggered traps as a function of time.

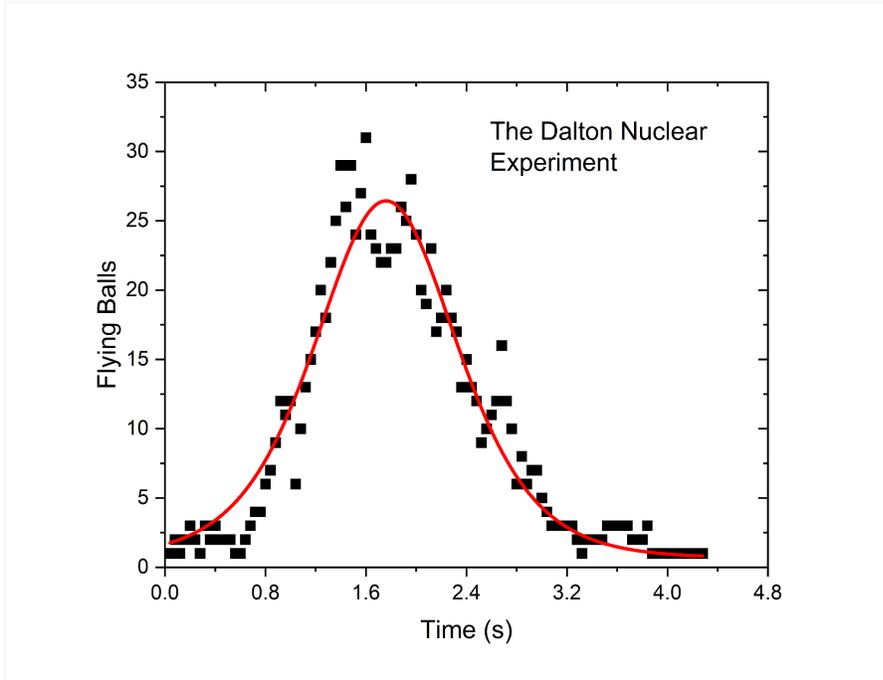

Figure 3, the DaltonNuclear experiment

3.3 *The Florence Experiment*. To measure the parameters for the Florence experiment we used two cameras, both commercial cell phones, one recording from above, the other from a side. Typical results for a single run are shown in Figure 4. Here, the number of flying balls has been fitted with the derivative of a logistic, while the number of untriggered traps with a simple logistic decay.

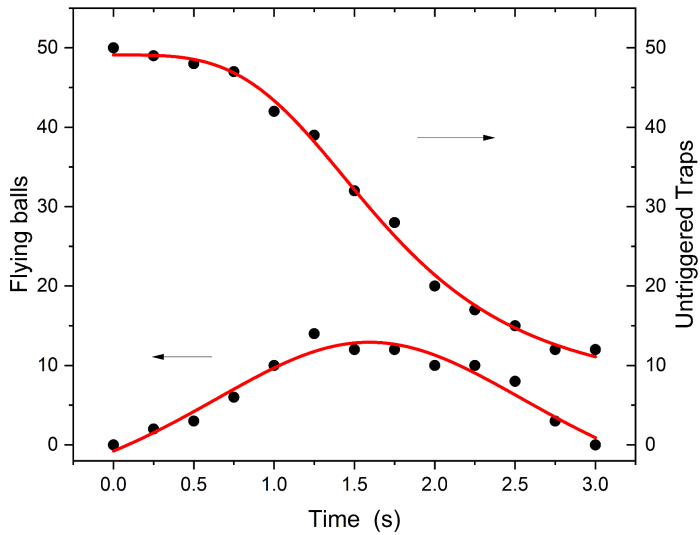

Fig. 4. Typical results of the Florence Experiments.

These experiments were well repeatable with the same setup even though, as it would be expected, the results of the measurements varied depending on random factors associated with the limited number of traps. In all tests, we found that the number of flying balls followed a bell-shaped curve.

## 4. Models

The derivative of a logistic and a logistic decay provide a first understanding of a typical phenomenon of growth and stabilization in complex systems such as in chemistry and in biology. Nevertheless, these functions do not provide a direct link to the physical properties of the system. For this purpose, we developed a mechanistic model that directly describes the experimental parameters using the methods of system dynamics.

We started from the observation that the system has a thermodynamic component: the potential energy stored in the springs of the mousetraps tends to be dissipated when the trap snaps. Balls are the catalyser that makes it possible for this energy to be released, being accumulated for a short while as kinetic energy of the flying balls and, ultimately, dispersed as high entropy thermal energy. This kind of phenomenon is common in biology and it describes a trophic chain, where an initial stock of metabolic energy is dissipated in steps by a series of predator-prey relationships. The simplest model that describes trophic chains is the well known Lotka-Volterra (LV) model (8) .

For the mousetrap experiment, we used a modification of the LV model that we call the "Single-Cycle Lotka Volterra" (SCLV) to take into account that the traps are not recharged (do not "reproduce") during the experiment. We developed it in previous studies to describe socio-economic systems that exploit non-renewable, or slowly renewable resources (2), (9). The SCLV model uses the same equations of the LV model, but the term that describes the growth of the prey population is set to zero. It is described by two coupled differential equations:

$dL_1/dt = - k_1 L_1 L_2$

$dL_2/dt = k_2 L_1 L_2 - k_3 L_2$

Here "$L_1$" ("Level 1) stands for the potential energy stored in the traps. It is measured using the number of cocked traps as a proxy. "$L_2$" ("Level 2") is the kinetic energy stored in the flying balls, measured in terms of the number of balls in the air. The $k_s$ are constants whose value depends on the unit of measurement of the levels.

The first equation of the model assumes that the number of traps snapping per unit time is proportional to both the number of flying balls and the number of untriggered traps. Of course, when either one of these parameters is zero, the reaction does not take place. The second equation describes the number of flying balls, also proportional to the number of flying balls and the number of cocked traps. Finally, the second term of the second equation describes the loss of energy of the flying ball as they eventually come to rest on the table.

The model was tested using a standard fitting procedure implemented in MATLAB. The fitted data were obtained from three experimental runs, all in the same set-up. The measurements for the flying balls were averaged, while those for the traps were also linearly interpolated since the measurements were taken at a different frame rate. As a result, the points do not correspond to integers. The results are shown in the figure.

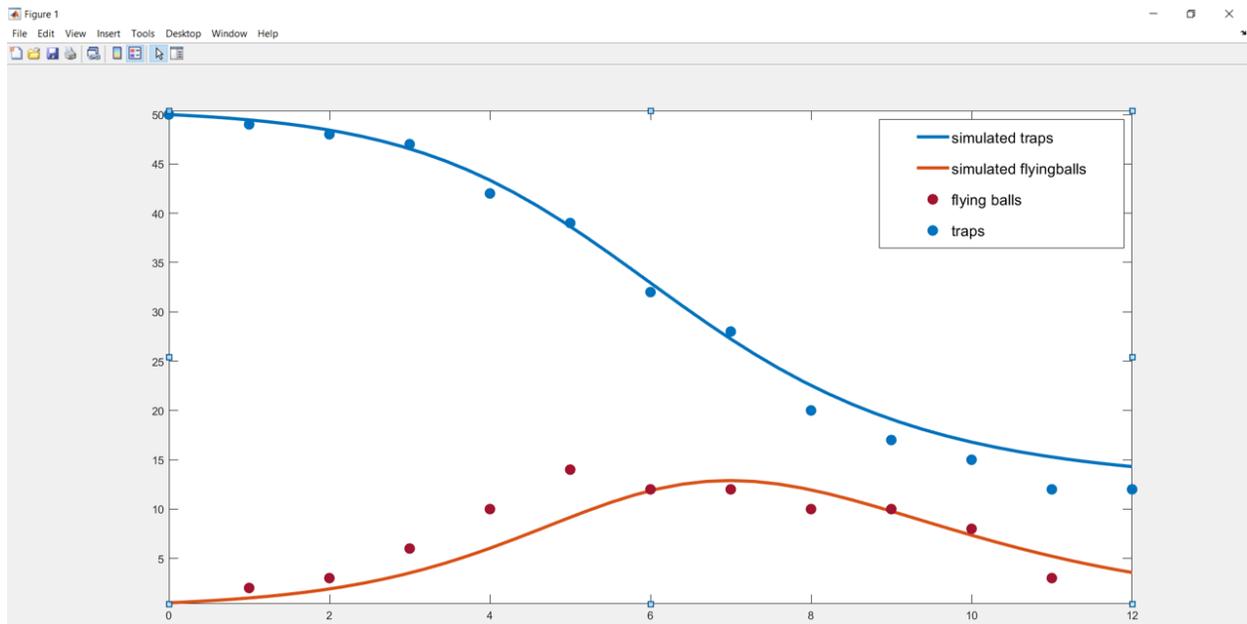

Figure 5. Results of the fitting using the SCLV model. The Y scale is the number of balls/untriggered traps. The X scale is the time in seconds x 4.

The coefficients for this specific fitting were found to be:

k1=0.0151
k2=0.303
k3=0.827

Remarkably, the fitting converged to a ratio of $k_2/k_1$ nearly equal to 2, which is what would be expected since every ball that triggers a trap releases two balls. This ratio corresponds to the "reproduction rate" in biological populations. Other experimental runs provided similar results, although in some cases the ratio was slightly higher than 2. It may have been the result of a ball triggering more than a single trap, or just fluctuations in the data.

The numerical coefficients can be used to determine the "net reproduction rate" according to the common definition in population biology. It is the number of new individuals (flying balls) divided by the number of deaths (balls on the ground). In the mousetrap experiment this parameter is given by the new balls $k_2L_1L_2$ per unit time divided by the fallen balls for the same unit time, that is $k_3L_2$. The result is $(k_2/k_3)L_1$. At t=0 (50 untriggered traps), this number is 1.8, close to 2, as one would expect. For t>0, the net reproduction rate falls and becomes smaller than one when the population of flying balls does not grow anymore. In this specific run, it occurs at the peak of the ball population peak, for $L_1$ = 28 untriggered traps. The final value of the net reproduction rate is 0.43 for 12 untriggered traps when there are no more flying balls.

We can also use these values to estimate the "critical mass" of the trap array. For $L_1$ <25, the net reproduction rate is smaller than one and the system should not show a chain reaction. We verified this result experimentally, finding that, indeed, for less than 25 traps the chain reaction either does not start or it involves only 2-3 traps. The variance is the result of the stochastic nature of the experimental setup. Note that, indeed, no experimental setup shown on the Web appears to use less than 25 traps. Apparently, this lower limit was found by trial and error.

## 5. Discussion

The mousetrap experiment was developed at the time of the "atomic age," in the late 1940s (1) , as a tool for explaining to the public the mechanism of the chain reaction that led to the release of energy in nuclear explosions. It was supposed to be noisy and spectacular and it succeeded at attracting much attention, In terms of simulating a nuclear reaction, the mousetrap experiment does a good job as a tool for the dissemination of scientific knowledge. Of course, the number of traps that can be reasonably used is much smaller than the number of nuclei involved in a nuclear chain reaction. There are also other differences, for instance the mousetrap experiment is usually set inside a box, whereas in the real world there does not exist a neutron reflector with comparable properties. Also, in most setups, balls and traps fly together and that is not the way fissionating nuclei behave. But, overall, the chain reaction effect is clearly shown in the experiment. It may well be that this experiment is the first public presentation of the effect we call today "enhancing" or "positive" feedback.

Using a controlled experimental set-up and well known system dynamics methods, we found that the experiment can be modeled and the model used to extract the fundamental parameters of the system. Among other things, we can use the model to find the "critical mass" of the mousetrap array, something that seems to have been found only by trial and error in previous experiments.

Our idea in re-examining this old experiment was not so much to add details to something already well-known. It was to show how general the phenomenon of energy transfer in trophic chains is, to the point that it can be simulated by mechanical devices such as mousetraps. The mousetrap array can be seen as a fully connected network that reacts to an external stimulus by processing and amplifying a signal. In this view, the first ball dropped on the mousetrap array can be seen as a packet of energy, a signal that is picked up by one of the mousetraps.The network amplifies the signal, in this specific system by about a factor of 30. That is, starting from one ball, there is a moment during the reaction when 30 traps are flying. In this sense, the system behaves like an electron multiplier of the kind used in night-vision systems. In that case, though, the gain may go from $10^5$ to $10^8$ and more.

Examples of this enhanced feedback behavior can be easily found in ecosystems. For instance if the balls can be seen as pathogens and the traps as susceptible people. In this case, the model can describe the flaring of an epidemic in a susceptible population according to the "SIR" (susceptible, infected, removed) model (10). It is well known that epidemics tend to flare and then subside when the "herd immunity" is reached.

Economic systems often behave as biological systems: for instance, if the ball is seen as a stock of exergy, then the traps can be seen as oil wells and the chain reaction involves re-investing part of this energy into exploiting new fossil resources. The "production" curve, in this case, is the well known "Hubbert curve" (11). In general, if the balls are seen as non-renewable or slowly renewable resources, the chain reaction describes the phenomenon of overexploitation. We observed this behavior also in fisheries (3), (9).

Among other things, note how what we called "net reproduction rate" in the previous section corresponds to the "$R_t$" rate in epidemiology and to the EROEI, or EROI (energy return for energy invested) (12) for several energy production systems that exploit non renewable resources, e.g. oil, as we noted in a previous study (13).

Perhaps the most interesting interpretation of the experiment is the behavior of the World Wide Web. In this case, the signal takes the name of "meme" and the resulting diffusion over the web is termed "going viral." In a previous paper, Bardi et al. showed that the same model used to fit the mousetrap experiment in the present paper could describe the diffusion of memes in the Web (14). The mousetrap experiment visually shows how rapidly these mechanisms create a true memetic explosion that then fades as rapidly as it has started. It is a warning of how the decisions taken at the macroscale are influenced by the amplification of signals that may be spurious and mislead people who act on the basis of what they can perceive.

Finally, note the similarity of the mousetrap experiment with another dynamic system based on a chain reaction among the elements of a network. It is the "Sandglass Model" or "Hourglass Model" proposed for the first time by Bak et al. (15). In this model, a grain of sand that falls on a previously existing pile generates an avalanche that propagates as the result of an enhancing feedback. Unlike the mousetrap system, the sandglass continues to generate avalanches as long as new grains of sand arrive from the upper reservoir. Nevertheless, the mousetrap system can approximately describe single avalanches in the sandglass system. The difference is that in the sandglass network the nodes are connected only to their nearest neighbors, whereas the mousetrap network is fully connected. The difference is small when the number of nodes is small, and single avalanches in a sandglass follow similar curves (16).

## 5. Conclusion

Mousetraps are the only simple mechanical device that can be bought at a hardware store that can be used to show a chain reaction, just like a sandglass is the only simple mechanical device that can show self-organized criticality. We don't know why these phenomena are so rare in hardware stores, but they are surely common in ecosystems and in other kinds of complex adaptive systems. We believe that the results we reported in this paper can be helpful to understand these systems, if nothing else to illustrate how chain reactions can easily go out of control and not only when they occur in a critical mass of fissile uranium.

**Acknowledgement**: the authors thank James Little, the author of the Berkeley mousetrap experiment and the authors of the Dalton Nuclear mousetrap experiments.

**Conflict of Interest.** The authors report no conflict of interest. This study was created using generic research funds.